\documentclass{amsart}

\usepackage{amsmath,amssymb,amsthm}

\newtheorem{theorem}{Theorem}
\newtheorem{lemma}[theorem]{Lemma}

\usepackage{tikz}
\usepackage{float}

\usepackage[authoryear,round]{natbib}
\usepackage{doi}

\newtheorem{corollary}[theorem]{Corollary}
\theoremstyle{definition}
\newtheorem{definition}[theorem]{Definition}
\newtheorem{conjecture}[theorem]{Conjecture}
\theoremstyle{remark}

\tikzset{
    tri/.style={black,line width=0.45pt},
    full/.style={black!15}
}

\newcommand{\fulltri}[3]{
    \fill[full] (#1)--(#2)--(#3)--cycle;
}

\newcommand{\drawtri}[3]{
    \draw[tri] (#1)--(#2)--(#3)--cycle;
}

\title[$\mathsf{Y}$ \& $\mathsf{X}$: Intermediate Logics]{$\mathsf{Y} \& \mathsf{X}$: Intermediate Logics of Convex Hulls and Cross-Polytopes}

\author{Yumin Ji}
\author{Han Xiao}

\address{The Tsinghua-UvA JRC for Logic, Tsinghua University, Beijing, China}

\email[Yumin Ji]{m2e07@naver.com}
\email[Han Xiao]{sheropen@163.com}

\begin{document}

\begin{abstract}
We investigate two logics of geometric Kripke frames: $\mathsf{Y}$, defined by finite convex hulls, and $\mathsf{X}$, defined by cross-polytopes. We characterize the validity of the Kreisel--Putnam axiom on convex hull frames by means of convex independence.
Our main technical results concern several reductions obtained using two translations, $C$ and $D$. In particular, we show that $\mathsf{X}$ is invariant under $C$, while $\mathsf{Y}$ is invariant under $D$. Moreover, the same translation $D$ reduces Medvedev logic to every logic in the interval $[\mathsf{Y}',\mathsf{Med}]$, where $\mathsf{Y}'$ is a naturally defined intermediate logic strictly between $\mathsf{Y}$ and $\mathsf{Med}$. Together with geometric arguments, these reductions establish that $\mathsf{Y}$, $\mathsf{X}$, and $\mathsf{Cheq}$ are pairwise incomparable, while each lies strictly between $\mathsf{PL}$ and Medvedev logic. We also prove that every logic in $[\mathsf{Y},\mathsf{Med}]$ is not finitely axiomatizable and that this interval contains continuum many distinct logics. In addition, every logic in $[\mathsf{Y}',\mathsf{Med}]$ is undecidable.
We prove that $\mathsf{X}$ is undecidable and $\mathsf{Y}\in \Pi^0_1$. In particular, we show that it is decidable whether a finite poset is isomorphic to a finite convex hull frame, and that the logics determined by configurations of a bounded number of points are decidable.
Finally, we extend our study to infinite convex hull frames. We show that inclusion yields $\mathsf{KC}$, whereas reverse inclusion yields $\mathsf{IPC}$. For the latter result, we use the Cantor set to construct $p$-morphisms onto arbitrary finite rooted frames.

\end{abstract}

\maketitle
\section{Introduction}
This project was initiated by a conceptual consideration of the first author. In particular, it began with an exploration of how logical tools could be used to study a geometric representation of concepts—namely, the theory of conceptual spaces proposed by \citet{Gardenfors2000}. The logic $\mathsf{Y}$ was discovered in the course of this conceptual endeavor. Results regarding the geometric counterparts of frames, i.e., that the logic defined by frames of infinite convex hulls ordered by inclusion is $\mathsf{KC}$ and that the logic defined by reversed inclusion is $\mathsf{IPC}$, are likewise conceptually motivated. Then, to investigate the relationships between these conceptually motivated logics, we employ the reduction technique introduced in \citet{xiao2026non}. Furthermore, we are interested in what these logics say about our conceptual representation—for instance, how difficult it is to learn a concept in a particular way under certain conceptual assumptions. Decidability results have thus come into the picture.

To illustrate this, we briefly introduce the theory of conceptual spaces proposed by  \citet{Gardenfors2000}. It is a geometric framework for conceptual representation in which objects are represented along cognitively meaningful quality dimensions and similarity is captured geometrically. A conceptual space is a metric space whose dimensions represent qualities. A color space, for example, may be structured by three quality dimensions: hue, saturation, and lightness. Each point represents a possible combination of quality values that a concrete instance may exhibit, while natural properties are represented by convex regions. Thus, “blue” may correspond to a convex region defined by certain ranges of hue, saturation, and lightness. A conceptual space of a concept is constructed as the Cartesian product of the conceptual spaces associated with the properties relevant to that concept.

We have considered two perspectives regarding concept learning under this geometric representation. One assumes that a correct way of characterizing a concept exists. That is, given a concept, the relevant qualities are fixed. Hence, to learn a concept is to identify its boundary. The other view does not assume that the correctness of a concept comes from its innate structure. Under this view, to learn a concept is to be able to recognize the distinctions and similarities between instances, and hence to become sufficiently acquainted with the diversity of its instances and the relationships among them. Then, three stages of conceptual assumptions are considered. First, we require that the representation be achieved with the minimum amount of information. Considering concept learning under this assumption leads to two different intermediate logics: $\mathsf{KC}$ and $\mathsf{Med}$. Next, we weaken the assumption: we keep the learning process finite, yet the representation is not necessarily achieved with the minimum amount of information. Under this weaker assumption, the two perspectives on learning lead to the logics $\mathsf{KC}$ and the proposed logic $\mathsf{Y}$, respectively. Finally, we weaken the assumption further by allowing the learning process to be infinite. This leads to $\mathsf{KC}$ and $\mathsf{IPC}$, respectively.

This paper presents the original technical results of this project.\footnote{For a detailed discussion of the conceptual considerations outlined above, please contact the first author for a draft.}

\section{Preliminary}
Throughout this paper, we assume familiarity with the basic theory of intermediate logics. 
We write $\mathsf{IPC}$ for \emph{intuitionistic propositional calculus}. 
By an (intuitionistic Kripke) frame we mean a poset. 

Given a frame $\mathbf{F}$, we write $\mathsf{Log}(\mathbf{F})$ for the propositional logic consisting of all formulas valid on the frame $\mathbf{F}$; if
$\mathsf{L} \subseteq \mathsf{Log}(\mathbf{F})$,
we call $\mathbf{F}$ an \emph{$\mathsf{L}$-frame}. If $\mathcal{C}$ is a class of frames, we write $\mathsf{Log}(\mathcal{C})$ for the intermediate logic consisting of all formulas valid on every frame of $\mathcal{C}$; in this case, we say that $\mathcal{C}$ \emph{characterizes} the logic.

\subsection{Posets as Kripke frames} 

Let $\mathbf{F} = \langle F,\leq \rangle$ be a finite poset; we write $x<y$ if $x\leq y$ and $x\neq y$. An element $x \in F$ is called \emph{maximal} if there is no $y \in F$ such that $x < y$; it is called a \emph{top} if everything is below it. A top is necessarily unique.
We say that $\mathbf{F}$ is a frame \emph{with a top} if it contains a top.
If $x<y$, we call $y$ an \emph{immediate successor} of $x$ if there is no $z$ such that $x<z<y$. 

The notation $x{\uparrow}$ (resp.\ $x{\downarrow}$) denotes the \emph{upset} $\{y\in F : x\leq y\}$ (resp.\ the \emph{downset} $\{y\in F : x\geq y\}$). A subframe $\mathbf{S}=\langle S, \leq \rangle$ of $\mathbf{F} = \langle F,\leq \rangle$ is called a \emph{generated subframe} if $S$ is upward closed in $F$; we say that $\mathbf{S}$ is \emph{generated by $x$} if $S = x{\uparrow}$. If $\mathbf{F}$ is generated by $x$, then $\mathbf{F}$ is called \emph{rooted} and $x$ the \emph{root} of $\mathbf{F}$.

Let $\mathbf{F}=\langle F,\le_0\rangle$ and 
$\mathbf{G}=\langle G,\le_1\rangle$ be posets. 
Their \emph{product} is the poset 
$\mathbf{F}\times\mathbf{G}
:= \langle F\times G,\le\rangle$, 
where the order $\le$ is defined coordinatewise by $(x,y)\le (x',y')$ if and only if $x\le_0 x'$ and $y\le_1 y'$.
For a poset $\mathbf{F}$, we write 
$\mathbf{F}^2 := \mathbf{F}\times\mathbf{F}$ and define inductively
$\mathbf{F}^n := \mathbf{F}^{\,n-1}\times\mathbf{F}$. 

\subsection{Morphisms} If $\mathbf{F}$ and $\mathbf{G}$ are posets and $f:F\to G$ is a map, we call $f$ a \emph{p-morphism} if for all $x,y\in F$, if $x\leq y$, then $f(x)\leq f(y)$ and
if $f(x)\leq u$, then there is some $y\geq x$ such that $f(y) = u$. If $f$ is onto, then we call $\mathbf{G}$ a \emph{p-morphic image} of $\mathbf{F}$.

\begin{theorem}[Folklore]
If $\mathbf{G}$ is a $p$-morphic image of $\mathbf{F}$, then $\mathbf{F}\vDash \varphi \text{ implies } \mathbf{G}\vDash \varphi$, for any formula $\varphi$.
\end{theorem}
 
\begin{theorem}[Jankov-de Jongh]
	Given a finite rooted frame $\mathbf{F}$, there exists a formula $\chi(\mathbf{F})$ such that for every frame $\mathbf{G}$, $$\mathbf{G} \nvDash \chi(\mathbf{F})
\quad\text{iff}\quad
\mathbf{F}
\text{ is a $p$-morphic image of a generated subframe of }
\mathbf{G}.$$The formula $\chi(\mathbf{F})$ is called the Jankov-de Jongh formula of $\mathbf{F}$.
\end{theorem}

\begin{corollary}[Folklore]
Given a class of finite frames $\textbf{C}$ that is closed under rooted generated subframes, for any finite rooted frame $\mathcal{F}$,
\[
\mathcal{F}\vDash \mathsf{Log}(\textbf{C}) \text{ iff } \mathcal{F} \text{ is a \emph{p}-morphic image of an element of } \textbf{C}.
\]
\end{corollary}

\subsection{Friedman translation}
\begin{definition}
Let $\rho$ be an arbitrary formula. For any formula $\varphi$, the \emph{Friedman translation} of $\varphi$ by $\rho$, denoted by $\varphi^{\rho}$, is defined inductively as follows:
\begin{itemize}
    \item $\varphi^{\rho} := \varphi \vee \rho$, if $\varphi$ is atomic;
    \item $(\varphi_0 \circ \varphi_1)^{\rho} := \varphi_0^{\rho} \circ \varphi_1^{\rho}$ for $\circ \in \{\wedge,\vee,\to\}$.
\end{itemize}
\end{definition}

\subsection{Medvedev logic}

In 1932, \cite{Kol32} suggested a constructive interpretation of intuitionistic logic as a calculus of problems. To make precise the description proposed by Kolmogorov, \cite{Med62} established the foundational framework for the logic of finite problems. We recall the definition of the \emph{Medvedev frame} and the corresponding \emph{Medvedev logic}.
\begin{definition}
For $n \ge 1$, the \emph{Medvedev frame on $n$ atoms} $\mathbf{P}_0(n)$ is defined by
$\mathbf{P}_0(n) 
:= \bigl\langle 
\{ X \subseteq [n] : X \neq \varnothing \}, 
\supseteq 
\bigr\rangle$, where $[n]:=\{1,2,\ldots,n\}$.
The \emph{Medvedev logic} $\mathsf{Med}$ is the intermediate logic of all Medvedev frames, that is, $\mathsf{Med}:= \mathsf{Log}\bigl(\{ \mathbf{P}_0(n): n \geq 1 \}\bigr)$.
\end{definition}
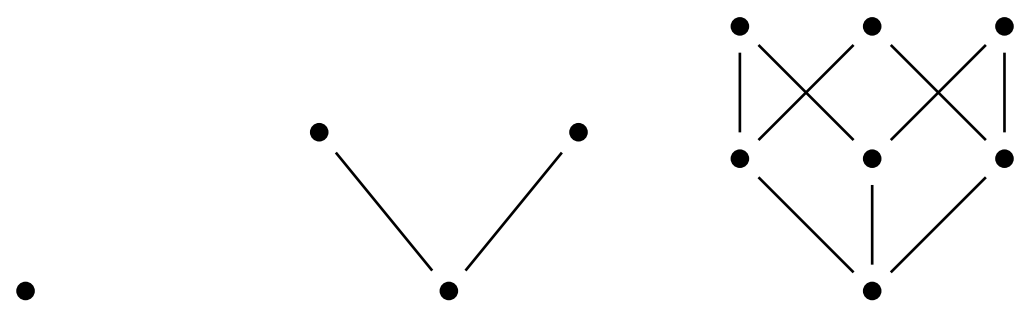
\begin{figure}[H]
\centering
\begin{tikzpicture}[
scale=0.35,
every node/.style={
    circle,
    fill=black,
    minimum size=3.5pt,
    inner sep=0pt
},
edge/.style={
    line width=0.5pt,
    shorten >=3pt,
    shorten <=3pt
}
]

\begin{scope}[xshift=0cm] 
    \node (a1) at (0,1) {};
    
\end{scope}

\begin{scope}[xshift=8cm] 
    \node (b1) at (0,1) {};
    \node (b2) at (-2.45,4) {};
    \node (b3) at (2.45,4) {};

    \draw[edge] (b1) -- (b2);
    \draw[edge] (b1) -- (b3);
\end{scope}

\begin{scope}[xshift=16cm] 
    \node (c1) at (0,1) {};
    \node (c2) at (-2.5,3.5) {};
    \node (c3) at (0,3.5) {};
    \node (c4) at (2.5,3.5) {};
    \node (c5) at (-2.5,6) {};
    \node (c6) at (0,6) {};
    \node (c7) at (2.5,6) {};

    \draw[edge] (c1) -- (c2);
    \draw[edge] (c1) -- (c3);
    \draw[edge] (c1) -- (c4);
    \draw[edge] (c2) -- (c5);
    \draw[edge] (c2) -- (c6);
    \draw[edge] (c3) -- (c5);
    \draw[edge] (c3) -- (c7);
    \draw[edge] (c4) -- (c6);
    \draw[edge] (c4) -- (c7);
\end{scope}

\end{tikzpicture}

\caption{Medvedev frames.}
\label{fig:med}
\end{figure}
\begin{lemma}[Maksimova, Skvortsov, \& Shehtman]
If $\mathbf{F}$ is a finite rooted frame with a top, then $\mathbf{F}$ is a $\mathsf{Med}$-frame.		
\label{lem:topmedframe}
\end{lemma}
Inheriting such a tradition on the frame for logic of finite problems, the order in our \emph{finite Boolean algebra} here is reverse inclusion $\supseteq$. 
\begin{definition}
For $n \ge 1$, the \emph{Boolean algebra on $n$ generators} $\mathbf{P}(n)$ is defined by $\mathbf{P}(n):=\bigl\langle 
\mathcal{P}([n]), \supseteq \bigr\rangle$, where $\mathcal{P}([n])$ denotes the powerset of $[n]$.
\end{definition}

\begin{definition}
The intermediate logic $\mathsf{KC}$ is obtained from $\mathsf{IPC}$ by adding the weak law of excluded middle: $\neg p\vee\neg\neg p$. It is also known as \textit{Jankov's logic} or the \textit{logic of weak excluded middle}.
\end{definition}

\begin{theorem}\label{KC-fin}
    $\mathsf{KC}=\bigcap_{n\geq 1}\mathsf{Log}\bigl(\mathbf P_0(n)^{\mathrm{op}}\bigr)$.
\end{theorem}

\begin{proof}
It follows from Corollary 1 of \cite{MaksimovaSkvortsovShehtman1979}.
\end{proof}

\subsection{The logic $\mathsf{Cheq}$}

In 2003, the \emph{modal logic of chequered subsets} $\mathsf{L}_{\infty}$ was introduced by \cite{vBBG03}.   
This logic plays an important role in the study of modal logics of 
various topological spaces; see, for example, \cite{vBB07, AvBB03}.
Beyond its role in the non-classical logics, $\mathsf{L}_{\infty}$ also arises in the study of \emph{modal logics of forcing}; see, for example, \cite{xiao2024modal}.
\cite{litak2018some} later denoted its 
intermediate analogue by $\mathsf{Cheq}$.
Whether $\mathsf{L}_{\infty}$ or $\mathsf{Cheq}$ is finitely axiomatizable is a longstanding open problem. This question was answered in the negative by \cite{xiao2026non}.
\begin{definition}
Let $\mathbf{V}$ be the \emph{two-fork}, that is, the poset consisting 
of three elements $\{w_0,w_1,w_2\}$ such that $w_0$ is the root and $w_1$ and $w_2$ 
are incomparable maximal elements. Define recursively 
$\mathbf{V}_1 := \mathbf{V}$ and 
$\mathbf{V}_{n+1} := \mathbf{V}_n \times \mathbf{V}$,
where $\times$ denotes the usual product of posets. 
Every point $x$ of $\mathbf{V}_n$ can be associated with an $n$-tuple $(x_0, x_1, \ldots, x_{n-1})$, where $x_i \in \{w_0, w_1, w_2\}$. 
The \emph{logic $\mathsf{Cheq}$} is the intermediate logic characterized by 
the class $\{\mathbf{V}_n : n \geq 1\}$.    
\end{definition}
\begin{figure}[H]
\centering
\begin{tikzpicture}[
scale=0.35,
every node/.style={
    circle,
    fill=black,
    minimum size=3.5pt,
    inner sep=0pt
},
every label/.style={
    circle=none,
    fill=none,
    inner sep=1pt
},
edge/.style={
    line width=0.5pt,
    shorten >=3pt,
    shorten <=3pt
}
]

\begin{scope}[xshift=0cm]
    \node (a0)  at (0,1) {};
    \node (a1)  at (-2.45,4) {};
    \node (a2)  at (2.45,4) {};

    \draw[edge] (a0) -- (a1);
    \draw[edge] (a0) -- (a2);
\end{scope}

\begin{scope}[xshift=10cm]
    \node (b00)  at (0,1) {};

    \node (b01)  at (-3,3.5) {};
    \node (b10)  at (-1,3.5) {};
    \node (b20)  at (1,3.5) {};
    \node (b02)  at (3,3.5) {};

    \node (b11)  at (-3,6) {};
    \node (b21)  at (-1,6) {};
    \node (b12)  at (1,6) {};
    \node (b22)  at (3,6) {};

    \draw[edge] (b00) -- (b01);
    \draw[edge] (b00) -- (b10);
    \draw[edge] (b00) -- (b20);
    \draw[edge] (b00) -- (b02);

    \draw[edge] (b01) -- (b11);
    \draw[edge] (b01) -- (b21);

    \draw[edge] (b10) -- (b11);
    \draw[edge] (b10) -- (b12);

    \draw[edge] (b20) -- (b21);
    \draw[edge] (b20) -- (b22);

    \draw[edge] (b02) -- (b12);
    \draw[edge] (b02) -- (b22);
\end{scope}

\end{tikzpicture}

\caption{$\mathbf{V}_1$ and $\mathbf{V}_2$.}
\label{fig:C1C2}
\end{figure}
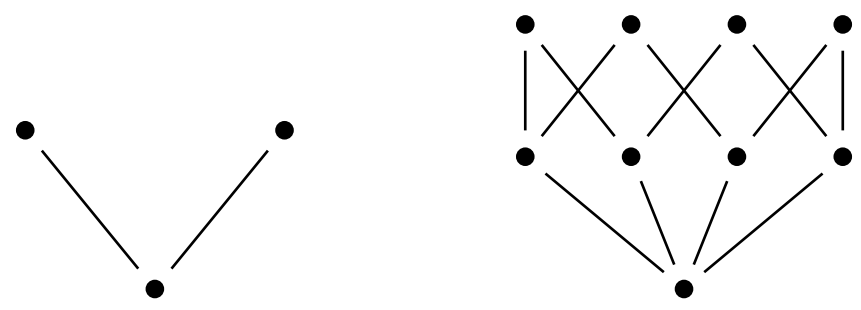
 
\subsection{Simplices}

\begin{definition}
Let $S=\{p_0,p_1,\ldots,p_k\} \subseteq \mathbb{R}^n$. An \textit{affine combination} is $x=\Sigma_{i=0}^k \lambda_ip_i$ where $\lambda_i \in \mathbb{R}$ and $\Sigma_{i=0}^k\lambda_i = 1$.
It is a \textit{convex combination} if $\lambda_i\geq0$ for all $i$. The set of all convex combinations is the \textit{convex hull} of $S$. 
The set $S$ is \textit{affinely independent} if no point in $S$ is an affine combination of the others. 
A \textit{$k$-simplex} is the convex hull of $k+1$ affinely independent points, and the points are its \textit{vertices}.
\end{definition}

\begin{figure}[H]
\centering
\begin{minipage}[t]{0.22\textwidth}
\centering
\begin{tikzpicture}[scale=0.8]
  \coordinate (A) at (0,0);

  \fill (A) circle (1.6pt);
  \node[below=4pt] at (A) {$a$};
\end{tikzpicture}

\end{minipage}
\hfill
\begin{minipage}[t]{0.22\textwidth}
\centering
\begin{tikzpicture}[scale=1]
  \coordinate (A) at (0,0);
  \coordinate (B) at (1.6,0);

  \draw[tri] (A)--(B);
  \fill (A) circle (1.6pt);
  \fill (B) circle (1.6pt);

  \node[below=4pt] at (A) {$a$};
  \node[below=4pt] at (B) {$b$};
\end{tikzpicture}

\end{minipage}
\hfill
\begin{minipage}[t]{0.22\textwidth}
\centering
\begin{tikzpicture}[scale=1]
  \coordinate (A) at (1.7,0.2);
  \coordinate (B) at (0.6,1.5);
  \coordinate (C) at (0,0);

  \fulltri{A}{B}{C}
  \drawtri{A}{B}{C}

  \fill (A) circle (1.6pt);
  \fill (B) circle (1.6pt);
  \fill (C) circle (1.6pt);

  \node[right=3pt] at (A) {$a$};
  \node[above=2pt] at (B) {$b$};
  \node[left=3pt] at (C) {$c$};
\end{tikzpicture}

\end{minipage}
\hfill
\begin{minipage}[t]{0.22\textwidth}
\centering
\begin{tikzpicture}[scale=1]
  \coordinate (A) at (1.85,0.45);
  \coordinate (B) at (0.95,1.55);
  \coordinate (C) at (0,0.95);
  \coordinate (D) at (0.35,0);

  \fill[full] (B)--(C)--(D)--cycle;
  \fill[full] (B)--(D)--(A)--cycle;

  \draw[tri] (B)--(C);
  \draw[tri] (C)--(D);
  \draw[tri] (D)--(A);
  \draw[tri] (A)--(B);
  \draw[tri] (B)--(D);

  \draw[black!65,dotted,line width=0.45pt] (C)--(A);

  \fill (A) circle (1.6pt);
  \fill (B) circle (1.6pt);
  \fill (C) circle (1.6pt);
  \fill (D) circle (1.6pt);

  \node[right=3pt] at (A) {$a$};
  \node[above=2pt] at (B) {$b$};
  \node[left=3pt] at (C) {$c$};
  \node[below=3pt] at (D) {$d$};
\end{tikzpicture}

\end{minipage}
\caption{$n$-simplices}
\end{figure}
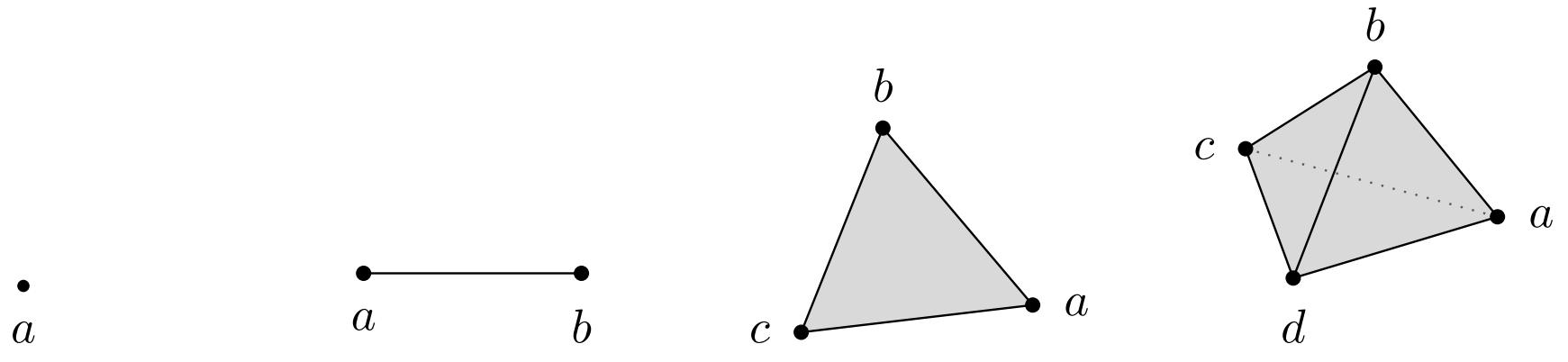

Observe that, for any subset of a simplex's vertices, the convex hull of the points in that subset is itself a simplex. For ease of notation, we include the relevant definition below; such simplices are called faces of the original simplex.

\begin{definition}
    Let $\sigma$ be a $k$-simplex defined by $S=\{p_0,p_1,\ldots,p_k\}$. A simplex $\tau$ defined by $T \subseteq S$ is a \emph{face} of $\sigma$.
\end{definition}

\section{The new logics: Y \& X}\label{5}
We consider the class of the dual frames of convex hulls as defined in Definition~\ref{def:dualconvexhull}. By Theorem~\ref{thm:Yop}, the logic valid on all these frames is $\mathsf{KC}$ as well. 

\begin{definition}
    Given a nonempty set $S=\{p_0,p_1,\ldots,p_k\} \subseteq \mathbb{R}^n$, define the \emph{dual frame of finite convex hulls} by $\mathbf{F}^{\mathrm{op}}_{S}:=\langle \{\mathrm{conv}(X): \emptyset\neq X\subseteq S \}, \subseteq \rangle$. The \emph{dual finite convex hull logic} $\mathsf{Y}^{\mathrm{op}}$ is the intermediate logic of all dual frames $\mathbf{F}^{\mathrm{op}}_{S}$. 
\label{def:dualconvexhull}
\end{definition}
\begin{theorem}
$\mathsf{Y}^{\mathrm{op}}=\mathsf{KC}$.
\label{thm:Yop}
\end{theorem}

\begin{definition}
Let $S=\{p_0,p_1,\ldots,p_k\} \subseteq \mathbb{R}^n$. The \emph{frame of finite convex hulls of} $S$ is $\mathbf{F}_{S}:=\langle \{\mathrm{conv}(X): \emptyset\neq X\subseteq S \}, \supseteq \rangle$. The \emph{logic of finite convex hulls in} $\mathbb{R}^n$, denoted by $\mathsf{Y}_n$, is the intermediate logic of all convex hull frames of all nonempty finite point sets $S\subseteq \mathbb{R}^n$, that is, $\mathsf{Y}_n:= \bigcap_{S\subseteq \mathbb{R}^n} \mathsf{Log}(\mathbf{F}_S)$. The \emph{logic of finite convex hulls} $\mathsf{Y}$ is the intermediate logic of all convex hull frames, that is, $\mathsf{Y}:= \bigcap_{n\geq 1} \mathsf{Y}_n$.
\label{def:logic_Y}
\end{definition}

In what follows, we present several properties of this logic. In this paper, we primarily focus on those that reveal a connection with geometry. We begin with a geometric interpretation of the \emph{Kreisel-Putnam axiom} $\mathsf{KP}:= (\lnot p \to (q \lor r)) \to ((\lnot p \to q)\lor (\lnot p \to r))$. To our knowledge, only one conceptual interpretation of this axiom has previously been discussed in \cite{holliday2017modal}. The interpretation developed here provides a second one, with a simple geometric correspondence.

\begin{theorem}
The following statements are equivalent:   
\begin{enumerate}
    \item $\mathbf{F}_S \vDash \mathsf{KP}$.
    \item $S$ is convexly independent.
    \item $\mathbf{F}_S \cong \mathbf{P}_0(\lvert S \rvert)$. 
\end{enumerate}
\label{thm:grassmann}
\end{theorem}
\begin{proof}
    (1. $\Rightarrow$ 2.) Assume that $\mathbf{F}_S \vDash \mathsf{KP}$ and that $S$ is not convexly independent. Let $a$ be a point in $S$ such that $a\in \mathrm{conv}(S\setminus \{a\})$. We now give a valuation $V$ on $\mathbf{F}_S$. For the propositional variable $p$, let $V(p)=\{a\}$ in $\mathbf{F}_S$. 
Every point $x\neq a$ is contained in some convex hull that does not contain $a$ and is maximal with respect to inclusion. Therefore, if there is only one such convex hull, then it contains all points in $S\setminus \{a\}$, and so contains $ \mathrm{conv}(S\setminus \{a\})$. Hence $a\notin \mathrm{conv}(S\setminus \{a\})$, a contradiction! Thus there are at least two distinct maximal convex hulls that do not contain $a$. We then divide those maximal convex hulls into two non-empty families $F_0$ and $F_1$. Let $V(q)= \bigcup_{C\in F_0} C\!\uparrow$ and $V(r)= \bigcup_{C\in F_1} C\!\uparrow$. Let $\mathcal{M}=(\mathbf{F}_S, V)$.
For any convex hull $C$, $C\vDash_{\mathcal{M}} \lnot p$ iff $a\notin C$. Thus, if $C\vDash_{\mathcal{M}} \lnot p$, then $C \in V(q)$ or $C \in V(r)$ and so $C\vDash_{\mathcal{M}} q\lor r$. Hence $\mathrm{conv}(S)\vDash_{\mathcal{M}} \lnot p\to (q\lor r)$. For any $C\in F_1$, because it does not contain $a$, $C\vDash_{\mathcal{M}} \lnot p$. Since $C$ is maximal and belongs to $F_1$, $C \notin V(q)$, so $\mathrm{conv}(S)\nvDash_{\mathcal{M}} \lnot p \to q$. By the same argument, $\mathrm{conv}(S)\nvDash_{\mathcal{M}} \lnot p \to r$. Therefore, $\mathrm{conv}(S)\nvDash_{\mathcal{M}} \mathsf{KP}$ and so $\mathbf{F}_S \nvDash \mathsf{KP}$, a contradiction!

    (2. $\Rightarrow$ 3.) If $S$ is convexly independent, then $f(X)=\mathrm{conv}(X)$ is an isomorphism from $\mathbf{P}_0(\lvert S \rvert)$ to $\mathbf{F}_S$.

    (3. $\Rightarrow$ 1.) It is well-known that $\mathsf{KP} \in \mathsf{Med}$. Hence $\mathbf{F}_S \vDash \mathsf{KP}$.
\end{proof}
\begin{corollary}
    $\mathsf{KP}\notin \mathsf{Y}$.
\label{cor:kp}
\end{corollary}

Now applying Theorem~\ref{thm:grassmann}, we give a proof of Theorem~\ref{thm:Yop}.
\begin{proof}[Proof of Theorem~\ref{thm:Yop}]
Every frame $\mathbf{F}^{\mathrm{op}}_{S}$ has a top $\mathrm{conv}(S)$. Recall that $\mathsf{KC}$ is the logic
of all finite frames with a top. Then every such frame validates $\mathsf{KC}$. Hence $\mathsf{KC}\subseteq \mathsf{Y}^{\mathrm{op}}$. 

If $S$ is convexly independent, then by Theorem~\ref{thm:grassmann}, $\mathbf{F}^{\mathrm{op}}_{S}\cong \mathbf{P}_0(\lvert S \rvert)^{\mathrm{op}}$. As a conclusion, $\mathsf{Y}^{\mathrm{op}}\subseteq \bigcap_{n\geq 1} \mathsf{Log}(\mathbf{P}_0(n)^{\mathrm{op}})=\mathsf{KC}$. The two inclusions establish the desired result.   
\end{proof}

We now characterize another logic of rich geometric meaning, one that is also deeply connected to logics such as $\mathsf{Cheq}$, $\mathsf{Med}$, and $\mathsf{Y}$. The geometric description of $\mathsf{Cheq}$, namely, the face poset of an $n$-cube, motivates an analogous construction using cross-polytopes, the dual polytopes of $n$-cubes. Let $e_i$ denote the $i$-th unit vector in $\mathbb{R}^n$.
For any $n\geq 1$, let $U_n:=\mathrm{conv}(\{\pm e_1, \pm e_2,\ldots,\pm e_n\})$ be the $n$-dimensional cross-polytope. 
\begin{definition}
Let $\mathbf{X}_n$ denote the poset consisting of all nonempty faces of $U_n$, ordered by reverse inclusion. The \emph{logic of cross-polytopes} $\mathsf{X}$ is the intermediate logic of all $\mathbf{X}_n$, that is, $\mathsf{X}=\bigcap_{n\geq 1} \mathsf{Log}(\mathbf{X}_n)$.
\label{def:logic_X}
\end{definition}
Every nonempty proper face of $U_n$ is a simplex. The rooted subframe generated by a facet of $U_n$ is isomorphic to the Medvedev frame $\mathbf{P}_0(n)$.

\section{Where $\mathsf{Y}$ meets $\mathsf{X}$}
We now investigate $\mathsf{Y}$ and $\mathsf{X}$ in deep, examining their positions and properties. Although $\mathsf{Y}$ and $\mathsf{X}$ seem structurally quite different, they are remarkably unified by Medvedev logic. Specifically, in both $\mathsf{X}$ and the related series of logics for $\mathsf{Y}$ (such as $\mathsf{Y}'$), the proper rooted generated subframes are exactly Medvedev frames. This enables the uniform application of reduction $D$ to them.

\subsection{Comparison of related logics}

We turn to its position among intermediate logics that can be characterized in terms of geometric objects. The starting point is a structural meaning of $\mathsf{PL}$ on frames, provided by \cite{adam2023intermediate}.

\begin{lemma}[Adam-Day, Bezhanishvili, Gabelaia, \& Marra]
Let $\mathbf{F}$ be a finite poset. Then $\mathbf{F}\vDash \mathsf{PL}$ iff the following conditions hold.
\begin{enumerate}
    \item If $d(x)=2$, then $\lvert \{y\in \mathbf{F}:y>x\} \rvert \leq 2$.
    \item If $d(x)>2$, then the set $\{y\in \mathbf{F}:y>x\}$ is connected.
\end{enumerate} 
\label{lem:PL0}
\end{lemma}

\begin{lemma}
    Let $S$ be a finite set and let $W \subseteq \mathcal{P}(S)$. Assume that $\emptyset \notin W$, and that  $\{x\}\in W$ for every $x\in S$. For any $w\in W$, define a graph $G_w:=(w, E_w)$, where $E_w=\{\{x,y\}:x\neq y\in w,\{x,y\}\in W\}$. Then $\langle W, \supseteq \rangle\vDash \mathsf{PL}$ iff for any $w\in W$, $G_w$ is connected.
\label{lem:PL1}
\end{lemma}
\begin{proof}
($\Rightarrow$) Assume that $\langle W, \supseteq \rangle\vDash \mathsf{PL}$ and that there exists a $w'\in W$ with $\lvert w' \rvert$ smallest and $G_{w'}$ is disconnected. If $\lvert w \rvert=2$, then $G_w$ is connected and so $\lvert w' \rvert\geq 3$. The graph $G_{w'}$ is the disjoint union of its connected components. For any $v\in W$ and $v\subsetneq w'$, $G_v$ is connected and thus there exists exactly one connected component such that $G_v$ is contained in it. Consequently, $\{w\in \mathbf{F}:w>w'\}$ is disconnected. Therefore, this contradicts the conditions of Lemma~\ref{lem:PL0}. 

($\Leftarrow$) For any $w\in W$, if $\lvert w \rvert=2$, then $w$ has exactly two immediate successors. If  $\lvert w \rvert\geq 3$, then $d(w)>2$. Take any $v_0,v_1\in w\!\uparrow \setminus\{w\}$, and choose $x\in v_0$ and $y\in v_1$. Since $G_w$ is connected, there is a path between $x$ and $y$ in $G_w$. The edges of this path give a path between $v_0$ and $v_1$. Thus, $w\!\uparrow\setminus \{w\}$ is connected. By Lemma~\ref{lem:PL0}, $\langle W, \supseteq \rangle\vDash \mathsf{PL}$. 
\end{proof}

We take a geometric approach to give a proof of the following theorem.
\begin{theorem}
    $\mathsf{PL} \subseteq \mathsf{Cheq}\cap \mathsf{Y}\cap \mathsf{X}$.
\label{thm:PL}
\end{theorem}
\begin{proof}
    In the case of $\mathsf{Cheq}$, \cite{xiao2024modal} proved that every frame $\mathbf{V}_n$ is isomorphic to the poset of nonempty faces of the $n$-cube, ordered by reverse inclusion. In the case of $\mathsf{X}$, the frames characterizing $\mathsf{X}$ are similarly obtained from the cross-polytope. Let $P$ be one of those polytopes. For any nonempty face $F$ of $P$, define
    \[
    f(F)=\{F_0: F_0 \text{ is a vertex of } F\}.
    \]
Each face is the convex hull of its vertices, thus $F\subseteq F'$ iff $f(F)\subseteq f(F')$. The poset of nonempty faces of $P$ is isomorphic to $\langle \{f(F): F \text{ is a nonempty face of } P\}, \supseteq \rangle$. 

Thus, $G_{f(F)}$ is the graph whose vertices are the $0$-faces of $F$ and whose edges are the $1$-faces of $F$, and is the $1$-skeleton of $F$. Since the $1$-skeleton of every nonempty face of a convex polytope is connected, by Lemma~\ref{lem:PL1}, every such frame validates $\mathsf{PL}$ and so $\mathsf{PL} \subseteq \mathsf{Cheq}\cap \mathsf{X}$. 

In the case of $\mathsf{Y}$, let $S$ be a finite nonempty set in $\mathbb{R}^n$. For any nonempty subset $X$ of $S$, define
\[
f'\left(\mathrm{conv}(X)\right)=\mathrm{conv}(X)\cap S.
\]
It gives an isomorphism from $\mathbf{F}_S$ to $\mathbf{F}'_S:=\langle \{X \subseteq S: \emptyset\neq X= \mathrm{conv}(X) \cap S\}, \supseteq \rangle$, with inverse $f'^{-1}(X)=\mathrm{conv}(X)$. For any $X\in \mathbf{F}_S$, we claim that $G_{X}$ is connected.

For any $x\neq y\in G_{X}$, $x$ and $y$ are adjacent iff $\{x,y\}\in \mathbf{F}'_S$ iff $[x,y]\cap S=\{x,y\}$. If $x\neq y\in X$, then $[x,y]\subseteq \mathrm{conv}(X)$ and so $[x,y]\cap S\subseteq \mathrm{conv}(X) \cap S=X$. Since $S$ is finite, we can then enumerate all points of $[x,y]\cap S$ in the order from $x$ to $y$: $p_0, p_1,\ldots, p_s$, where $p_0=x,p_s=y$ and $[p_i,p_{i+1}]\cap S=\{p_i,p_{i+1}\}$. Thus, $p_i$ and $p_{i+1}$ are adjacent and so give a path from $x$ to $y$ in $G_X$. Thus, $G_{X}$ is connected. Therefore, every frame $\mathbf{F}'_S$ and $\mathbf{F}_S$ validates $\mathsf{PL}$ and so $\mathsf{PL}\subseteq \mathsf{Y}$. 
\end{proof}
Recall that in \cite{xiao2026non}, we construct a reduction from $\mathsf{Med}$ to $\mathsf{Cheq}$. For any formula $\varphi$, let $p, q$ be two distinct propositional variables that do not occur in $\varphi$. Define $\psi_0(p)$ and $\psi_1(p,q)$ as follows:
\[\psi_0 (p):= p \vee \neg p \quad \psi_1 (p,q):=\neg p \vee \neg q.	
\]

Then we define a formula $C(\varphi): = (\varphi^{\psi_0}\to \psi_1)\to  \psi_1$.

Now we give another reduction from $\mathsf{Med}$ to $\mathsf{X}$. For any formula $\varphi$, let $p$ be a propositional variable that do not occur in $\varphi$. We define a formula $D(\varphi):=p \lor (p\to \varphi)$.

The \emph{G\"{o}del--Dummett logic}, also known as \emph{chain logic}, is defined by $\mathsf{LC}:= \mathsf{IPC}+\mathsf{GD}$, where $\mathsf{GD}:=(p\to q)\lor (q\to p)$ is the \emph{G\"{o}del--Dummett axiom}. The following theorem establishes a reduction from $\mathsf{LC}$ to $\mathsf{Y}$ by the translation $C$. 

\begin{theorem}
    For any formula $\varphi$, $\varphi \in \mathsf{LC}$ iff $C(\varphi) \in \mathsf{Y}$.
\label{thm:LCY}
\end{theorem}
\begin{proof}
($\Rightarrow$) Suppose $C(\varphi)\notin \mathsf{Y}$, then there is a model $\mathcal{M}$ on some frame $\mathbf{F}_S$ and there is convex hull $C_0$ in $\mathbf{F}_S$, such that $C_0\vDash_{\mathcal{M}} \varphi^{\psi_0}\to \psi_1$ and $C_0 \nvDash_{\mathcal{M}} \psi_1$. For all convex hulls contained in $C_0$, choose the minimal one under the inclusion with refuting $\psi_1$, let it be $C_1$, then $C_1 \vDash_{\mathcal{M}} \varphi^{\psi_0}\to \psi_1$ and $C_1 \nvDash_{\mathcal{M}} \psi_1$. So $C_1 \nvDash_{\mathcal{M}} \varphi^{\psi_0}$. Since  $C_1 \nvDash_{\mathcal{M}} \psi_1$, then there are points $x, y \in C_1 \cap S$, $\{x\}\vDash_{\mathcal{M}} p$ and $\{y\}\vDash_{\mathcal{M}} q$. If $x=y$, then $\{x\}\vDash_{\mathcal{M}} \psi_0$ and so $\{x\}\vDash_{\mathcal{M}} \varphi^{\psi_0}$. Because $\{x\}\nvDash_{\mathcal{M}} \psi_1$, we obtain that $\{x\}\nvDash_{\mathcal{M}} \varphi^{\psi_0}\to\psi_1$, a contradiction. Consequently, $x\neq y$. Since $[x,y]=\mathrm{conv}({x,y})\subseteq C_1$ refutes $\psi_1$, then $C_1=[x,y]$ by the minimality. If there is another $z\neq x\in [x,y] \cap S$ with $\{z\}\vDash_{\mathcal{M}}p$, then $z\neq y$ and $[z,y]\nvDash_{\mathcal{M}}\psi_1$. Because $[z,y]\subsetneq [x,y]$, it is a contradiction. Thus, $x$ is the only point in $[x,y]$ validating $p$ and $y$ is the only point validating $q$. Because $S$ is finite, we can enumerate all points in $[x,y]\cap S$, let it be $\{p_0,p_1,\ldots,p_s\}$ with $p_0=x$ and $p_1=y$. We now only consider the counter-model restricted to $[x,y]\!\uparrow$. Since $[p_i,p_j]\vDash_{\mathcal{M}} \lnot p$ iff $x\notin [p_i,p_j]$, then the convex hulls refuting $\psi_0=p\lor \lnot p$ are exactly a chain: $[x,y]\supsetneq [x, p_{s-1}]\supsetneq\ldots\supsetneq [x,p_1]$. By the Pruning Lemma, $[x,y]\nvDash_{\mathcal{M}} \varphi^{\psi_0}$ implies $[x,y] \nvDash_{\mathcal{M}^{\psi_0}} \varphi$, which means $\varphi$ is refuted in a model of a finite chain. Therefore, $\varphi \notin \mathsf{LC}$.

($\Leftarrow$) Suppose $\varphi \notin \mathsf{LC}$, let $\mathcal{M}_0$ be a counter-model in an $s$-chain whose root refutes $\varphi$, with valuation $V_0$. Now we choose $s+1$ collinear points $p_0<p_1<\ldots<p_s$. In the frame $\mathbf{F}_{\{p_0,p_1,\ldots,p_s\}}$, let $V_1$ be a valuation on it such that $V_1(p)=p_0$ and $V_1(q)=p_s$. Let $\mathcal{M}_1=\langle \mathbf{F}_{\{p_0,p_1,\ldots,p_s\}}, V_1\rangle$, in this model, the points refuting $\psi_0=p\lor \lnot p$ are exactly the $s$-chain $[p_0,p_s]\supsetneq [p_0, p_{s-1}]\supsetneq\ldots\supsetneq [p_0,p_1]$. Let $\mathcal{M}_2=\langle \mathbf{F}_{\{p_0,p_1,\ldots,p_s\}}, V_2\rangle$, where $V_2(p)=V_1(p)$, $V_2(q)=V_1(q)$, and $V_2(r)=V_0(r)$ for any propositional variable $r$ occurring in $\varphi$. By the Pruning Lemma, $[p_0,p_s] \nvDash_{\mathcal{M}_2} \varphi^{\psi_0}$ since $[p_0,p_s] \nvDash_{\mathcal{M}_2^{\psi_0}} \varphi$. Because only $\{p_0\}\vDash_{\mathcal{M}_2} p$ and $\{p_s\}\vDash_{\mathcal{M}_2} q$, thus $[p_0,p_s] \nvDash_{\mathcal{M}_2} \psi_1$ and $C\vDash_{\mathcal{M}_2} \psi_1$ For any $C> [p_0,p_s]$. Consequently, $[p_0,p_s] \vDash_{\mathcal{M}_2} \varphi^{\psi_0}\to \psi_1$ and so $[p_0,p_s] \nvDash_{\mathcal{M}_2} (\varphi^{\psi_0}\to \psi_1)\to \psi_1$. Therefore, $C(\varphi)\notin \mathsf{Y}$.
\end{proof}

 The following theorems show that $\mathsf{X}$ is invariant under $C$ and $\mathsf{Y}$ is invariant under $D$.
\begin{theorem}
    For any formula $\varphi$, $\varphi \in \mathsf{X}$ iff $C(\varphi) \in \mathsf{X}$.
\label{thm:xxx}
\end{theorem}
\begin{proof}

($\Rightarrow$) Suppose $C(\varphi)\notin \mathsf{X}$. Then there is a model $\mathcal{M}$ on some $\mathbf{X}_n$, and a nonempty face $U$ of $U_n$, such that $U\vDash_{\mathcal{M}} \varphi^{\psi_0}\to \psi_1$ and $U\nvDash_{\mathcal{M}} \psi_1$.  For all non-empty faces contained in $U$, choose the minimal one under the inclusion with refuting $\psi_1$, let it be $U'$, then $U'\vDash_{\mathcal{M}} \varphi^{\psi_0}\to \psi_1$ and $U' \nvDash_{\mathcal{M}} \psi_1$. So $U' \nvDash_{\mathcal{M}} \varphi^{\psi_0}$. Since $U' \nvDash_{\mathcal{M}} \psi_1$, then there are points $x,y\in U'$, such that $\{x\}\vDash_{\mathcal{M}} p$ and  $\{y\}\vDash_{\mathcal{M}} q$. If $x=y$, then ${x}\vDash_{\mathcal{M}}\psi_0$ and so $\{x\}\vDash_{\mathcal{M}} \varphi^{\psi_0}$. Because $\{x\}\nvDash_{\mathcal{M}} \psi_1$, we obtain that $\{x\}\nvDash_{\mathcal{M}} \varphi^{\psi_0}\to\psi_1$, a contradiction. Consequently, $x\neq y$. If $U'$ is a proper face of $U_n$, then $U'$ is a simplex. Since the simplex $[x,y]=\mathrm{conv}({x,y})\subseteq U'$ refutes $\psi_1$, then $U'=[x,y]$ by the minimality. The same conclusion holds for $U'=U_1$. Thus, $x$ is the only point in $[x,y]$ validating $p$ and $y$ is the only point validating $\lnot p$. In this case, only a single point $[x,y]$ does not force $\psi_0$.
If $U'=U_n$ with $n\geq 2$, then by the minimality, $x$ and $y$ are antipodal points. For any other vertex $z$, regardless of whether $z\vDash_{\mathcal{M}}p$ or $z\vDash_{\mathcal{M}}q$, it forms, together with $x$ or $y$, a proper face satisfying $\psi_1$, a contradiction! Hence $x$ is the only point in $U_n$ validating $p$ and $y$ is the only point validating $q$. Therefore, for any face $U''$, $U''\vDash_{\mathcal{M}}\lnot p$ iff $x\notin U''$. Thus, if $\mathcal{M}^{\psi_1}=\langle W^{\psi_1}, \supseteq, V\rangle$, then $W^{\psi_1}=\{U_n\}\bigcup \{U: U \text{ is proper face of } U_n, x\in U\neq \{x\}\}$, and so $\langle W^{\psi_1},\supseteq \rangle\cong \mathbf{X}_{n-1}$. 
As a conclusion, $\mathcal{M}^{\psi_0}$ is a model on a single point or on $\mathbf{X}_{n-1}$. But by the Pruning Lemma, $U'\nvDash_{\mathcal{M}}\varphi^{\psi_0}$ iff $U'\nvDash_{\mathcal{M}^{\psi_0}} \varphi$, then $\varphi \notin \mathsf{X}$. 

($\Leftarrow$) Suppose $\varphi \notin \mathsf{X}$, let $\mathcal{M}_0$ be a counter-model in $\mathbf{X}_n$ whose root refutes $\varphi$, with valuation $V_0$. Now we choose $U_{n+1}$. In the frame $\mathbf{X}_{n+1}$, let $V_1$ be a valuation on it such that $V_1(p)=+e_0$ and $V_1(q)=-e_0$. Let $\mathcal{M}_1=\langle \mathbf{X}_{n+1}, V_1\rangle$, in this model, the points refuting $\psi_0=p\lor \lnot p$ form $\mathbf{X}_{n}$ by using the same argument. 
Let $\mathcal{M}_2=\langle \mathbf{X}_{n+1}, V_2\rangle$, where $V_2(p)=V_1(p)$, $V_2(q)=V_1(q)$, and $V_2(r)=V_0(r)$ for any propositional variable $r$ occurred in $\varphi$. 
By the Pruning Lemma, $U_{n+1} \nvDash_{\mathcal{M}_2} \varphi^{\psi_0}$ since $U_{n+1} \nvDash_{\mathcal{M}_2^{\psi_0}} \varphi$. 
Since $V_2(p)=+e_0$, $V_2(q)=-e_0$, then $U_{n+1} \nvDash_{\mathcal{M}_2} \psi_1$ and for any proper face $U$ of $U_{n+1}$, $U\vDash_{\mathcal{M}_2} \psi_1$. Consequently, $U_{n+1} \vDash_{\mathcal{M}_2} \varphi^{\psi_0}\to \psi_1$ and so $U_{n+1} \nvDash_{\mathcal{M}_2} (\varphi^{\psi_0}\to \psi_1)\to \psi_1$. Therefore, $C(\varphi)\notin \mathsf{X}$.
\end{proof}

\begin{theorem}
    For any formula $\varphi$, $\varphi \in \mathsf{Y}$ iff $D(\varphi) \in \mathsf{Y}$.
\label{thm:yyy}
\end{theorem}
\begin{proof}
($\Rightarrow$) It is obvious that $\left( \varphi \to D(\varphi) \right) \in \mathsf{IPC}$, thus if $\varphi \in \mathsf{Y}$, then $D(\varphi)\in \mathsf{Y}$.

($\Leftarrow$) If $\varphi \notin Y$, then there is a finite points set $S$ and a valuation $V$ on $\mathbf{F}_S$, such that $\text{conv}(S) \nvDash_{(\mathbf{F}_S, V)} \varphi$. Let $S'=S\cup\{s\}$, where $s\notin S$. Then the $\text{conv}(S)\!\uparrow$ in $\mathbf{F}_{S'}$ is exactly $\mathbf{F}_S$. Let the valuation of $s$ on $\mathbf{F}_{S'}$ be $\text{conv}(S)\!\uparrow$, the valuation of variables in $\varphi$ is equal to $V$. Then $\mathbf{F}_{S'}$ together with this valuation is a countermodel for $D(\varphi)$, thus $D(\varphi)\notin \mathsf{Y}$.
\end{proof}
Following the same argument in Theorem~\ref{thm:yyy}, we can prove that $\mathsf{Cheq}$ is invariant under $D$.
\begin{theorem}
    For any formula $\varphi$, $\varphi \in \mathsf{Cheq}$ iff $D(\varphi) \in \mathsf{Cheq}$.
\label{thm:ccc}
\end{theorem}

The following result builds a reduction from $\mathsf{Med}$ to $\mathsf{X}$ by the translation $D$. Proof will be provided
later.
\begin{theorem}
    For any formula $\varphi$, $\varphi \in \mathsf{Med}$ iff $D(\varphi) \in \mathsf{X}$.
\label{thm:medx}
\end{theorem}
\begin{proof}
    ($\Rightarrow$) For a formula $\varphi$, suppose that $D(\varphi) \notin \mathsf{X}$. There is a countermodel $\mathcal{M}$ on $\mathbf{X}_n$, a face $U$ of $\mathbf{X}_n$ refutes $D(\varphi)=p \lor (p \to \varphi)$. Then $U\nvDash_{\mathcal{M}} p$, and there exists $U'\subsetneq U$ such that $U' \vDash_{\mathcal{M}} p$ and $U' \nvDash_{\mathcal{M}} \varphi$. $U'$ is a simplex and so $U'\uparrow$ is a Medvedev frame. Hence $\varphi \notin \mathsf{Med}$.

($\Leftarrow$) Suppose $\varphi \notin \mathsf{Med}$, then there is a countermodel $\mathcal{M}'$ on $\mathbf{P}_0(n)$ such that its root refutes $\varphi$. Consider the cross-polytope $U_n$, its facet $U$ is an $n$-simplex. Now extending the $\mathcal{M}'$ from the rooted subframe generated by the facet $U$ to $\mathbf{X}_n$ by the valuation of $p$ is the non-empty faces of $U$. Thus $U\vDash_{\mathcal{M}'} p$ and $U\nvDash_{\mathcal{M}'} \varphi$. Since $U_n \nvDash_{\mathcal{M}'} p$, so $U_n \nvDash_{\mathcal{M}'} p \to \varphi$ and $U_n \nvDash_{\mathcal{M}'} p \lor (p \to \varphi)$.  Hence $D(\varphi) \notin \mathsf{X}$. 
\end{proof}

\begin{theorem}
    $\mathsf{PL}\subsetneq \mathsf{Y} \subsetneq \mathsf{Med}$.
\label{thm:whereisY}
\end{theorem}
\begin{proof}
By Theorem~\ref{thm:grassmann}, the logic $\bigcap_{S} \mathsf{Log}(\mathbf{F}_S)$ is the Medvedev logic $\mathsf{Med}$ for all $S$ which is convexly independent. Consequently, $\mathsf{Y}=\bigcap_{S} \mathsf{Log}(\mathbf{F}_S)\subseteq \mathsf{Med}$. Corollary~\ref{cor:kp} shows that $\mathsf{KP}\in \mathsf{Med}\setminus \mathsf{Y}$. Therefore, $\mathsf{Y} \subsetneq \mathsf{Med}$. 

Since $\mathsf{GD}\in \mathsf{LC}$, by Theorem~\ref{thm:LCY}, $C(\mathsf{GD})\in \mathsf{Y}$. 
In \cite{xiao2026non}, for any $\varphi$, $\varphi\in \mathsf{Med}$ iff $C(\varphi)\in \mathsf{Cheq}$. It is well-known that $\mathsf{GD}\notin \mathsf{Med}$, then $C(\mathsf{GD})\notin \mathsf{Cheq}$. By Theorem~\ref{thm:PL}, $C(\mathsf{GD})\notin \mathsf{PL}$ and $\mathsf{PL}\subseteq \mathsf{Y}$. As a conclusion, $\mathsf{PL}\subsetneq \mathsf{Y}$.
\end{proof}

\begin{theorem}
    $\mathsf{PL}\subsetneq \mathsf{X} \subsetneq \mathsf{Med}$.
\label{thm:whereisX}
\end{theorem}
\begin{proof}
For a given cross-polytope, choose one of its facets; it is a simplex. Thus, the rooted generated subframe above this facet is the nonempty faces poset of a simplex ordered by reverse inclusion. Therefore, $\mathsf{X} \subseteq \mathsf{Med}$. \cite{litak2018some} gives a counter-model of $\mathsf{KP}$ on the $2$-cube, which is also a cross-polytope, thus $\mathsf{KP}\notin \mathsf{X}$. Thus, $\mathsf{X} \subsetneq \mathsf{Med}$.
 
By theorem~\ref{thm:PL}, $\mathsf{PL}\subseteq \mathsf{X}$.
By theorem~\ref{thm:medx}, since $\mathsf{KP}\in \mathsf{Med}$, then $D(\mathsf{KP})\in \mathsf{X}$. It is not difficult to see that $D(\mathsf{KP})\notin \mathsf{PL}$. Therefore, $\mathsf{PL}\subsetneq \mathsf{X}.$
\end{proof}

\begin{theorem}
    $\mathsf{PL}\subsetneq \mathsf{Cheq} \subsetneq \mathsf{Med}$.
\label{thm:whereischeq}
\end{theorem}
\begin{proof}
By \cite{xiao2026non}, we have $\mathsf{Cheq} \subsetneq \mathsf{Med}$.

By Theorem~\ref{thm:PL}, $\mathsf{PL}\subseteq \mathsf{Cheq}$. On the other hand, by \cite{xiao2026non}, we have $C(\mathsf{KP})\in \mathsf{Cheq}$ since $\mathsf{KP}\in \mathsf{Med}$. By Theorem~\ref{thm:whereisX}, $\mathsf{KP}\notin \mathsf{X}$. Hence, by Theorem~\ref{thm:xxx}, $C(\mathsf{KP})\notin \mathsf{X}$, and therefore $C(\mathsf{KP})\notin \mathsf{PL}$. Thus, $\mathsf{PL}\subsetneq \mathsf{Cheq}$.
\end{proof}

\begin{lemma}
 $\mathsf{Cheq}\nsubseteq \mathsf{Y}$.
\label{lem:cheqnotincludedinY}
\end{lemma}
\begin{proof}
Let $\{p_0,p_1,p_2\}$ be three affinely independent points. They form a $2$-simplex. Let $x$ and $y$ be two points located in the interior of this simplex. Let $S=\{p_0,p_1,p_2,x,y\}$. We prove that, $\mathbf{F}_S$ is not a $p$-morphic image of any $\mathbf{V}_n$.

Suppose $f$ is a $p$-morphism from $\mathbf{V}_n$ to $\mathbf{F}_S$, without loss of generality, we choose the smallest $n$. We regard $\mathbf{V}_n$ as the poset of $n$-cube $Q_n$, see \cite{xiao2024modal,xiao2026non}. 

Choosing a linear function $P$, such that $P$ takes pairwise distinct values on the five points in $S$, and reordering $S$ by increasing order of the values of $P$, and denote the resulting sequence as $z_0<z_1<\ldots<z_4$. Let $X_i:=\text{conv}(\{z_0,z_1,\ldots,z_i\})$, then $\text{conv}(S)=X_4 \supsetneq X_3 \supsetneq \ldots \supsetneq X_0=z_0$ are a chain in $\mathbf{F}_S$. Since $f$ is a $p$-morphism, then there is a chain $Q_n=Y_4 \supsetneq Y_3 \supsetneq \ldots \supsetneq Y_0$ in $ Q_n$ such that $f(Y_i)=X_i$. Therefore, $n\geq 4$.

Since $f$ is a $p$-morphism, then there is a subface $Q_m$ of $Q_n$, such that $f(Q_m)=[p_0,p_1]$ and all vertices in $Q_m$ are sent to $p_0$ or $p_1$ by $f$. Furthermore, there is an edge $e_{0,1}$ in $Q_m$, wohse endpoints are sent to $p_0$ or $p_1$, respectively. Following the same argument, we obtain $e_{1,2}$ and $e_{0,2}$. If $n\geq 4$, then we can choose a coordinate direction $\vec{j}$ orthogonal to $e_{0,1}$, $e_{1,2}$, and $e_{0,2}$. Considering the two facets in $Q_n$ with normal vector $\vec{j}$. Then $e_{0,1}$, $e_{1,2}$, and $e_{0,2}$ belong to the two facets and so one facet $F$ contains two of those edges, thus $f(F)$ contains $p_0$, $p_1$ and $p_2$. Therefore, $f(F)=\text{conv}(\{p_0,p_1,p_2\})=S$, which contradicts the minimality of $Q_n$.   

Since $\mathbf{F}_S$ is not a $p$-morphic image of any $\mathbf{V}_n$, then $\mathbf{F}_S \nvDash \mathsf{Cheq}$ and so $\mathsf{Cheq}\nsubseteq \mathsf{Y}$. 
\end{proof}

\begin{theorem}
    $\mathsf{Cheq}$, $\mathsf{Y}$ and $\mathsf{X}$ are pairwise incomparable.
\label{thm:cheqyxincomparable}
\end{theorem}
\begin{proof}
Since $\mathsf{KP}\in \mathsf{Med}$, then $D(\mathsf{KP})\in \mathsf{X}$ (Theorem~\ref{thm:medx}). But $\mathsf{KP}\notin \mathsf{Y}$, then $D(\mathsf{KP})\notin \mathsf{Y}$(Theorem~\ref{thm:yyy}). Because $\mathsf{KP}\notin \mathsf{Cheq}$, then $D(\mathsf{KP})\notin \mathsf{Cheq}$ (Theorem~\ref{thm:ccc}). 

Since $\mathsf{GD}\in \mathsf{LC}$, then $C(\mathsf{GD})\in \mathsf{Y}$ (Theorem~\ref{thm:LCY}). But $\mathsf{GD}\notin \mathsf{X}$, then $C(\mathsf{GD})\notin \mathsf{X}$ (Theorem~\ref{thm:xxx}). Because $\mathsf{GD}\notin \mathsf{Med}$, then $C(\mathsf{GD})\notin \mathsf{Cheq}$ (see \cite{xiao2026non}).

Since $\mathsf{KP}\in \mathsf{Med}$, then $C(\mathsf{KP})\in \mathsf{Cheq}$ (see \cite{xiao2026non}). Because $\mathsf{KP}\in \mathsf{LC}$, then $C(\mathsf{KP})\in \mathsf{Y}$ (Theorem~\ref{thm:LCY}). But $\mathsf{KP}\notin \mathsf{X}$, then $C(\mathsf{KP})\notin \mathsf{X}$ (Theorem~\ref{thm:xxx}). 

Together with Lemma~\ref{lem:cheqnotincludedinY}, we obtain that $\mathsf{Cheq}$, $\mathsf{Y}$ and $\mathsf{X}$ are pairwise incomparable.
\end{proof}

Therefore, the positions of $\mathsf{Y}$ and $\mathsf{X}$ can be summarized as follows.
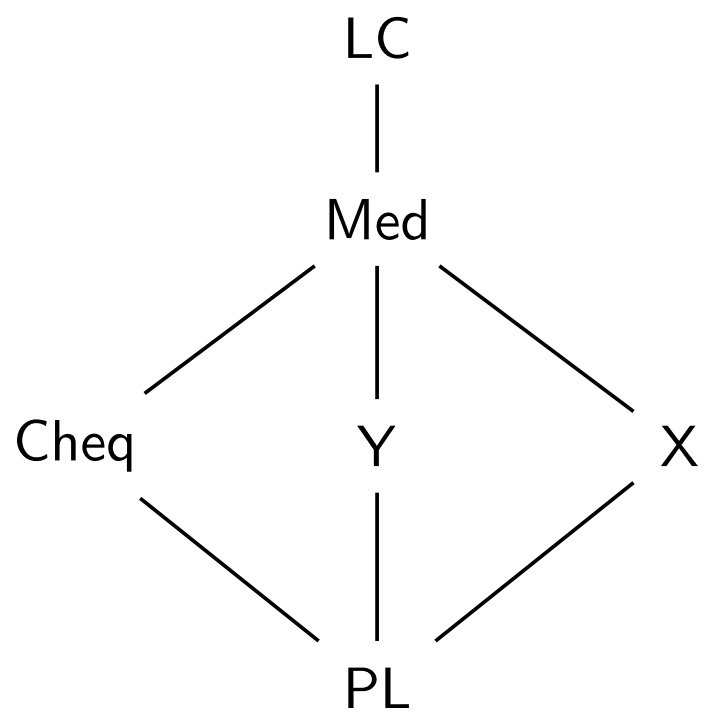
\begin{figure}[H]
    \centering
    \begin{tikzpicture}[
        line width=0.7pt,
        every node/.style={
            font=\large,
            inner sep=5pt,
            outer sep=0pt
        }
    ]
        \node (LC)   at ( 0,4.3) {$\mathsf{LC}$};
        \node (Med)  at ( 0,3.1) {$\mathsf{Med}$};
        \node (Cheq) at (-2,1.6) {$\mathsf{Cheq}$};
        \node (Y)    at ( 0,1.6) {$\mathsf{Y}$};
        \node (X)    at ( 2,1.6) {$\mathsf{X}$};
        \node (PL)   at ( 0,0)   {$\mathsf{PL}$};

        \draw (Med) -- (LC);
        \draw (PL) -- (Cheq) -- (Med);
        \draw (PL) -- (Y)    -- (Med);
        \draw (PL) -- (X)    -- (Med);
    \end{tikzpicture}
    \caption{Comparison of the relevant logics.}
\label{fig:logic-inclusions}
\end{figure}

\subsection{The non-finite axiomatizability and undecidability}

\begin{theorem}
The logic $\mathsf{X}$ is undecidable and not finitely axiomatizable.   
\label{thm:Xisundecidable}
\end{theorem}

\begin{proof}
    The undecidability of $\mathsf{Med}$ proved by AI in \cite{AK26,Paw26}, by Theorem~\ref{thm:medx}, $\mathsf{X}$ is undecidable. Together with the finite model property of $\mathsf{X}$, $\mathsf{X}$ is not finitely axiomatizable.
\end{proof}

\begin{theorem}
    $\mathsf{Y}$ is not finitely axiomatizable.
\end{theorem}
\begin{proof}
We will directly follow the proof of Claim~7 in \cite{fontaine2006ml}. The only modification for the convex hull setting of $\mathsf{Y}$ is a geometric generalization: we generalize the operation of taking the union of two subsets in the original proof to that of taking the convex hull of two sub-convex hulls, and we replace the cardinality of a set by the cardinality of the vertices of the minimal convex hull as a preimage. The rest of the argument can remain unchanged.
\end{proof}

\begin{theorem}
    For any logic $\mathsf{M}$ satisfying $\mathsf{Y}\subseteq\mathsf{M}\subseteq \mathsf{Med}$, $\mathsf{M}$ is not finitely axiomatizable.
\label{thm:M}
\end{theorem}
\begin{proof}
Suppose $\mathsf{M}$ is finitely axiomatizable with $s$ propositional variables. We directly adopt the structure $\mathbf{L}_{s,2^{s+3}}$, whose original construction is due to \cite{MaksimovaSkvortsovShehtman1979}. Since the proof that $\mathsf{Y}$ is not finitely axiomatizable employs exactly the  $\mathbf{L}_{s,2^{s+3}}$, we obtain that $\chi(\mathbf{L}_{s,2^{s+3}}) \in \mathsf{Y}$. For any $M \supseteq Y$, $\chi(\mathbf{L}_{s,2^{s+3}}) \in \mathsf{M}$.

Since every frame $\mathbf{L}'_{s,2^{s+3},m}\vDash \mathsf{Med}$ and $\mathsf{M}\subseteq \mathsf{Med}$, then  $\mathbf{L}'_{s,2^{s+3},m}\vDash \mathsf{M}$. Because $\mathsf{M}$ is finitely axiomatizable with $s$ propositional variables, we obtain $\mathbf{L}_{s,2^{s+3}}\vDash \mathsf{M}$ and $\mathbf{L}_{s,2^{s+3}}\vDash \chi(\mathbf{L}_{s,2^{s+3}})$, a contradiction!
\end{proof}

\begin{theorem}
$\mathsf{Y} \in \Pi^0_1$. 
\label{thm:YPi}
\end{theorem}
\begin{proof}

We follow the construction in Theorem~\ref{thm:PL}.
Let $S$ be a finite nonempty set in $\mathbb{R}^n$. For any nonempty subset $X$ of $S$,
$f'(\mathrm{conv}(X))=\mathrm{conv}(X)\cap S$ gives an isomorphism from $\mathbf{F}_S$ to $\mathbf{F}'_S:=\langle \{X \subseteq S: \emptyset\neq X= \mathrm{conv}(X) \cap S\}, \supseteq \rangle$. If we know which nonempty subset $X\subseteq S$ satisfies $X=\text{conv}(X)\cap S$, then we will know the frame $\mathbf{F}'_{S}$. Let $S=\{p_1,\ldots,p_n\}$. Then the set of points in $\mathbf{F}'_S$ can be coded as $W\subseteq \mathcal{P}([n])\setminus \{\emptyset\}$ satisfying: $X\in W$ iff $\{p_i:i\in X\}=\text{conv}(\{p_i: i\in X\})\cap S$. It gives the isomorphism:  $\langle W,\supseteq \rangle \cong \mathbf{F}'_S$. We assume that $S=\{p_1,\ldots,p_n\}\subseteq \mathbb{R}^{n-1}$, thereby we can associate each $p_i$ with its coordinates $p_i=(p_{i,1},p_{i,2},\ldots,p_{i,n-1})$. Since $\text{conv}(\{p_i:i\in X\})=\{\sum_{i\in  X}\lambda_i p_i : \lambda_i\geq 0, \sum_{i\in X} \lambda_i =1 \}$, then $p_j \in \text{conv}(\{p_i:i\in X\})$ iff there are $\{\lambda_i:i\in X\}$ such that $p_j=\sum_{i\in X} \lambda_i p_i$ with $\lambda_i\geq 0$ and $\sum_{i\in X} \lambda_i =1$. Let $\Psi_{j,X}(\bar p):= \exists(\lambda_i)_{i \in X}  \bigwedge_{m=1}^{n-1} (p_{j,m}=\sum_{i\in X} \lambda_ip_{i, m}) \bigwedge_{i\in X} \lambda_i \geq 0 \wedge \sum_{i\in X} \lambda_i =1$. Therefore, $p_j \in \text{conv}(\{p_i:i\in X\})$ can be expressed by $\Psi_{j,X}(\bar p)$, where $\bar p=(p_{1,1},p_{1,2},\ldots, p_{1,n-1}, p_{2,1},\ldots, p_{n,n-1})$.
Let $\Phi_{X}(\bar p):=\bigwedge_{j\in [n]\setminus X} \lnot \Psi_{j,X}(\bar p)$. If all $p_i$ are pairwise distinct, then $\Phi_{X}(\bar p)$ holds iff $\{p_i: i\in X\}=\text{conv}(\{p_i:i\in X\}) \cap S$.
Let $\Delta_n(\bar p):= \bigwedge_{1\leq i< j\leq n} \left(\sum_{m=1}^{n-1} (p_{i,m}-p_{j,m})^2>0\right)$. Then the points in $S$ are pairwise different iff $\Delta(\bar p)$ holds.

Let $\Omega_{n,W}:=\exists \bar p  \left( \Delta_n(\bar p) \land \bigwedge_{X\in W} \Phi_{X}(\bar p) \land \bigwedge_{X\notin W, \emptyset\neq X\subseteq [n]} \lnot  \Phi_{X}(\bar p) \right)$. There are pairwise distinct points $p_1,\ldots,p_n\in \mathbb{R}^{n-1}$, such that $g(X)=\{p_i:i\in X\}$ sends $W$ to all points in $\mathbf{F}'_S$, iff the formula $\Omega_{n,W}$ holds in $\mathbb{R}$.
There is an algorithm, for any $n$ and $W$ on input, it outputs $\Omega_{n,W}$ after finite steps, and $\Omega_{n,W}$ is a first-order sentence in the language of ordered fields with integer coefficients. By Tarski's Decision Theorem for real closed fields, it is decidable whether $\mathbb{R}\vDash \Omega_W$. 

For any fixed $n$, enumerate all $W\subseteq \mathcal{P}([n])\setminus \{\emptyset \}$ and remain those satisfying $\mathbb{R} \vDash \Omega_W$, we then obtain a finite, effectively computable list $T_n$ of frames $\langle W, \supseteq \rangle$. Each frame of $T_n$ is isomorphic to $\mathbf{F}'_S$ for some $S$ of $n$ points. Conversely, each $\mathbf{F}'_S$ with some $S$ of $n$ points is isomorphic to a frame on $T_n$. Consequently, $T_1, T_2,\ldots$ effectively enumerate all frames of finite convex hulls. 

We now prove that it is decidable whether a finite poset is isomorphic to some frame of finite convex hull. Let $\mathbf{F}$ be a finite poset with $n$ maximal points, then $F$ is isomorphic to some frame of finite convex hull, iff $F$ is isomorphic to some frame in $T_n$. Since the isomorphism between finite posets is decidable and $T_n$ is finite, effectively computable, the required procedure follows.

It is well-known that the validity of a propositional formula on a finite frame is decidable. Let $E(n, \ulcorner\varphi\urcorner)$ express that each frame of $T_n$ validates $\varphi$, where $\ulcorner\varphi\urcorner$ is the G\"odel code of $\varphi$. Then $E$ is a total computable predicate. It is easy to see that $\varphi\in \mathsf{Y}$ iff for any $n$, $E(n, \ulcorner\varphi\urcorner)$, thus $\mathsf{Y}\in \Pi^0_1$.
\end{proof}

The proof of Theorem~\ref{thm:YPi} also yields the following decidability results.
\begin{theorem} 
It is decidable whether a given finite poset is isomorphic to a finite convex hull frame.
\label{thm:posetdecidable}
\end{theorem}

The same construction also decides validity when the number of points is bounded.

\begin{theorem}
For each $n\geq 1$, let $\mathsf{Y}^{\leq n}:=  \bigcap_{1\leq |S|\leq n} \mathsf{Log}(\mathbf{F}_S)$. Then $\mathsf{Y}^{\leq n}$ is decidable.
\label{Y^ndecidable}
\end{theorem}

We obtained a sequence of decidable logics $\mathsf{Y}^{\leq 1}\supseteq \mathsf{Y}^{\leq 2}\supseteq \ldots $. It is obvious that $\mathsf{Y}=\bigcap_{n\geq 1} \mathsf{Y}^{\leq n}$. We have the following conjecture.
\begin{conjecture}
$\mathsf{Y}$ is undecidable.   
\end{conjecture}

Given $n\geq 2$, let $S_n=\{p_0,p_1,\ldots,p_n\}$ be an affinely independent set. Then its convex hull is an $n$-simplex; let $q_n$ be the barycenter of this simplex. Let $S'_n:=S_n\cup \{q_n\}$. 

\begin{definition}
 The \emph{convex hull frame of $S'_n$} is $\mathbf{F}_n:=\langle \{\text{conv}(X): \emptyset\neq X\subseteq S'_n \}, \supseteq \rangle$. Let $\mathsf{Y}_n':=\mathsf{Med} \cap \mathsf{Log}(\mathbf{F}_n)$, and let $\chi_n$ be the Jonkov-de Jongh formula of $\mathbf{F}_n$. The logic $\mathsf{Y}':=\bigcap_{n\geq 2} \mathsf{Y}'_n$. 
\end{definition}

\begin{lemma}
    $\mathsf{Y}\subsetneq \mathsf{Y}'\subsetneq \mathsf{Med}$.
\label{lem:whereisY'}
\end{lemma}
\begin{proof}
Since every frame $\mathbf{F}_n$ is a convex hull frame, thus $\mathsf{Y}\subseteq\mathsf{Y}_n'\subseteq \mathsf{Med}$, and so $\mathsf{Y}\subseteq\mathsf{Y}'\subseteq \mathsf{Med}$. 

By Theorem~\ref{thm:grassmann}, $\mathsf{KP}\notin \mathsf{Y}'$, because of the convex dependence of $S'_n$. We already have that $\mathsf{KP}\in \mathsf{Med}$.

Later we will see that $\varphi \in \mathsf{Med}$ iff $D(\varphi)\in \mathsf{Y}'$ in Theorem~\ref{thm:Y'Medreduction}. Hence $D(\mathsf{KP})\in \mathsf{Y}'$. Let $S=\{0,1,2,3\}$ be points set of real line. Then the rooted subframe generated by $[1,3]$ in $\mathsf{F}_{S}$ refutes $\mathsf{KP}$ since the convex dependence. Now let the valuation of this countermodel of $\mathsf{KP}$ extend to $\mathsf{F}_{S}$, and let $V(p):=[1,3]\uparrow$, $\mathcal{M}=(\mathbf{F}_S, V)$. Then $[0,3]\nvDash_{\mathcal{M}} p \lor (p\to \mathsf{KP})$. Therefore, $D(\mathsf{KP})\notin \mathsf{Y}$. Consequently, $\mathsf{Y}\subsetneq \mathsf{Y}'\subsetneq \mathsf{Med}$.
\end{proof}

\begin{theorem}
There are $2^{\aleph_0}$ many logics in the interval $[\mathsf{Y}', \mathsf{Med}]$
\label{thm:Y'Med}
\end{theorem}
\begin{proof}
We first prove that any $\mathbf{F}_m$ is not a $p$-morphic image of $\mathbf{F}_n$ when $n\neq m$. It is easy to see that $\mathbf{F}_n\cong \langle \mathcal{P}(S'_n)\setminus \{S_n, \emptyset\}, \supseteq \rangle$. Suppose $\mathbf{F}_m$ is a $p$-morphic image of $\mathbf{F}_n$ by the $p$-morphism $f$. Since the single point is the maximal point of the frame, the $p$-morphism $f$ introduces a map $f'$ from $S'_n$ to $S'_m$, such that $f(\{p_i\})=\{f'(p_i)\}$. It is easy to see that $f(X)=\{f'(p): p\in X\}$ for any points set $X$. Now consider the points set $S_m$, which is not a point in frame $\mathbf{F}_m$. Let $S$ be the set of $S'_n$ with $\{f'(p):p\in S\}=S_m$, suppose $S$ in a point in frame $\mathbf{F}_n$, then $f(S)=\{f'(p):p\in S\}=S_m$, but $S_m=f(S)$ is not a point in $\mathbf{F}_m$. Thus the nonempty set $S$ is not a point in $\mathbf{F}_n$, so $S=S_n$. Therefore, $\{f'(p):p\in S_n\}=S_m$ and $f'(q_n)=q_m$. Consequently, $n\geq m$. If $n>m$, then there exist two distinct points $P_0$ and $p_1$ such that $f'(p_0)=f'(p_1)$. Then $S_n\setminus \{p_0\}$ is a point in $\mathbf{F}_n$, but $\{f'(p):p\in S_n\setminus\{p_0\} \}=S_m$ and so $f(S_n\setminus \{p_0\})=S_m\notin \mathbf{F}_m$, a contradiction. Thus $n=m$.

Then we prove that $\mathbf{F}_m$ is not a $p$-morphic image of any rooted generated subframe of $\mathbf{F}_n$ when $n\neq m$. If this rooted generated subframe is not $\mathbf{F}_n$ itself, it would be a Medvedev frame. But any Medvedev frame satisfies $\mathsf{KP}$, while $\mathbf{F}_m$ doesn't. By Jonkov-de Jongh Theorem, $\mathbf{F}_n \vDash \chi_m$ iff $n\neq m$.

For any $A\subseteq\{2,3,\ldots\}$, let $\mathsf{L}_{A}:=\mathsf{Y}'+\{\chi_i:i\in A\}$, then $\mathsf{Y}' \subseteq \mathsf{L}_{A}\subseteq  \mathsf{Med}$ and $\chi_i \in \mathsf{L}_A$ iff $i\in A$. So we get at least $2^{\aleph_0}$ many distinct logics $\mathsf{L}_A$. Because there are at most $2^{\aleph_0}$ many distinct logics, thus the cardinality of logics in \mbox{$[\mathsf{Y}', \mathsf{Med}]$ is $2^{\aleph_0}$}.
\end{proof}

\begin{corollary}
 There are $2^{\aleph_0}$ many logics in the interval $[\mathsf{Y}, \mathsf{Med}]$, each of them is not finitely axiomatizable.
\label{cor:YMednotfa}
\end{corollary}

Using the reduction $D$, we can even prove that every logic in the entire interval is undecidable.

\begin{theorem}
 For any logic $\mathsf{M}$ satisfying $\mathsf{Y}' \subseteq \mathsf{M} \subseteq \mathsf{Med}$, for any formula $\varphi$, $\varphi \in \mathsf{Med}$ iff \mbox{$D(\varphi) \in \mathsf{M}$.}
\label{thm:Y'Medreduction}
\end{theorem}
\begin{proof}
Following the same argument in Theorem~\ref{thm:medx}, we obtain that, for any formula $\varphi$, $\varphi \in \mathsf{Med}$ iff $D(\varphi) \in \mathsf{Med}$ and  $\varphi \in \mathsf{Med}$ iff $D(\varphi) \in \mathsf{Y}'$.

Given a logic $\mathsf{M}$ in the interval $(\mathsf{Y}', \mathsf{Med})$, for any formula $\varphi$, $\varphi \in \mathsf{Med}$ implies $D(\varphi)\in \mathsf{Y}'$, thus $D(\varphi)\in \mathsf{M}$, and $\varphi \notin \mathsf{Med}$ implies $D(\varphi)\notin \mathsf{Med}$, thus $D(\varphi)\notin \mathsf{M}$. Consequently, $\varphi \in \mathsf{Med}$ iff $D(\varphi) \in \mathsf{M}$. 
\end{proof}
  
\begin{theorem}
 For any logic $\mathsf{M}$ satisfying $\mathsf{Y}' \subseteq \mathsf{M} \subseteq \mathsf{Med}$, $\mathsf{M}$ is undecidable.
\label{thm:Y'Medundecidable}
\end{theorem}
\begin{proof}
By Theorem~\ref{thm:Y'Medreduction}, $\varphi \in \mathsf{Med}$ iff $D(\varphi) \in \mathsf{M}$. The undecidability of $\mathsf{Med}$ implies that $\mathsf{M}$ is undecidable.
\end{proof}
We already proved that every logic of 
the interval $[\mathsf{Y}, \mathsf{Med}]$ is not finitely axiomatizable, and every logic of 
the interval $[\mathsf{Y}', \mathsf{Med}]$ is undecidable. We have the following conjecture:

\begin{conjecture}
For any logic $\mathsf{M}$ satisfying $\mathsf{Y}\subseteq \mathsf{M} \subseteq \mathsf{Med}$, $\mathsf{M}$ is undecidable.
\end{conjecture}

\section{The infinite case}
When $S$ is an arbitrary set in $\mathbb{R}^n$, we consider its convex hull as the smallest convex set containing $S$, that is, $\text{conv}(S)=\{\sum^{m}_{i=1}\lambda_i p_i : m\geq 1, p_i\in S, \lambda_i\geq 0 \text{ and } \sum^{m}_{i=1}\lambda_i=1\}=\bigcup_{\emptyset\neq X \subseteq_{\text{fin}} S} \text{conv}(X)$. If $S$ is an infinite set of points, we say it is \emph{convexly independent}, if for any $p\in S$, $p\notin \text{conv}(S\setminus \{p\})$.

\begin{definition}
Let $S\subseteq \mathbb{R}^n$ be an arbitrary, possibly infinite, set of points. The \emph{convex hull frame of} $S$ is $\mathbf{F}_{S}:=\langle \{\text{conv}(X): \emptyset\neq X\subseteq S \}, \supseteq \rangle$, and its \emph{dual frame} is $\mathbf{F}^{\mathrm{op}}_{S}:=\langle \{\text{conv}(X): \emptyset\neq X\subseteq S \}, \subseteq \rangle$. 
The \emph{logic of convex hulls} $\mathsf{Y}_{\infty}$ is the intermediate logic of all convex hull frames. The \emph{dual convex hull logic} $\mathsf{Y}_{\infty}^{\text{op}}$ is the intermediate logic of all dual convex hull frames.
\end{definition}

We now establish the infinite version of Theorem~\ref{thm:grassmann}.

\begin{theorem}
If $S$ is an infinite set of points. The following statements are equivalent:   
\begin{enumerate}
    \item $\mathbf{F}_S \vDash \mathsf{KP}$.
    \item $S$ is convexly independent.
    \item $\mathbf{F}_S$ is an infinity Skvortsov frame.
    \item $\mathsf{Log}(\mathbf{F}_S)=\mathsf{Skvo}$.
\end{enumerate}
\label{thm:grassmann_inf}
\end{theorem}
\begin{proof}
($1. \Rightarrow 2.$) 
Assume that $\mathbf{F}_S \vDash \mathsf{KP}$ and that $S$ is not convexly independent. Let $a$ be a point in $S$ such that $a\in \mathrm{conv}(S\setminus \{a\})$. By the definition, we can choose a finite set $X=\{p_0,p_1,\ldots,p_n\}$ which is minimal under inclusion, such that $X\subseteq S\setminus\{a\}$ and $a\in \text{conv}(X)$. Thus  $a\notin \text{conv}(X\setminus \{p_0\})$. We now give a valuation $V$ on $\mathbf{F}_S$. Let $V(p)=\{\{a\}\}$, $V(q)= \bigcup_{p_0\notin C} C\!\uparrow$ and $V(r)= \bigcup_{\{p_1,\ldots,p_n\}\nsubseteq C} C\!\uparrow$. Let $\mathcal{M}=(\mathbf{F}_S, V)$. For any convex hull $C$, $C\vDash_{\mathcal{M}} \lnot p$ iff $a\notin C$. 
If $a\notin C$, then $X\nsubseteq C$, thus $p_0\notin C$ or $\{p_1,\ldots,p_n\}\nsubseteq C$. Hence $C\vDash_{\mathcal{M}} q\lor r$ and $\text{conv}(X)\vDash_{\mathcal{M}}\lnot p \to (q\lor r)$. Since $p_0\notin V(q) $ and $\text{conv}\{p_1,\ldots,p_n\}\notin V(r)$, then $\text{Conv}(X)\nvDash_{\mathcal{M}}\lnot p \to q$ and $\text{Conv}(X)\nvDash_{\mathcal{M}}\lnot p \to r$. Putting everything together, $\text{Conv}(X)\nvDash_{\mathcal{M}} \mathsf{KP}$ and so 
$\mathbf{F}_S \nvDash \mathsf{KP}$, a contradiction!

($2. \Rightarrow 3.$) If $S$ is convexly independent, then for any $X\subseteq S$, $f(X)=\text{conv}(X)$ is an isomorphism from $\mathbf{F}_{S}$ to $\langle \mathcal{P}(S)\setminus \{\emptyset\}, \supseteq \rangle$.

($3. \Rightarrow 4.$)
\cite{Skv79} prove that $\mathsf{Skvo=\mathsf{Log}}\left(\langle \mathcal{P}(\omega)\setminus \{\emptyset\}, \supseteq \rangle\right)$. For any infinite $S$, it has a countable subset, the subframe generated by this subset is isomorphic to $\langle \mathcal{P}(\omega)\setminus \{\emptyset\}, \supseteq \rangle$, thus $\mathsf{Skov}\subseteq\mathsf{Log}(\mathbf{F}_S)\subseteq\mathsf{Log}\left(\langle \mathcal{P}(\omega)\setminus \{\emptyset\}, \supseteq \rangle \right)=\mathsf{Skov}$ and so $\mathsf{Log}(\mathbf{F}_S)=\mathsf{Skvo}$.

($4. \Rightarrow 1.$)
It is well-known that $\mathsf{KP} \in \mathsf{Skov}$. Hence $\mathbf{F}_S \vDash \mathsf{KP}$.
\end{proof}

\begin{theorem}
$\mathsf{Y}_{\infty}^{\text{op}}=\mathsf{KC}$.
\label{thm:Yinfopkc}
\end{theorem}
\begin{proof}
Because each frame $\mathbf{F}^{\text{op}}_S$ has a top $\text{conv}(S)$, thus $\mathsf{KC}\subseteq \mathsf{Y}_{\infty}^{\text{op}}$.

For any finite set $S_{\text{fin}}\subseteq \mathbb{R}^{n}$, let $S:=S_{\text{fin}} \times \mathbb{N}\subseteq \mathbb{R}^{n+1}$. The projection map $\pi(x_1,\ldots,x_n,x_{n+1})=(x_1,\ldots,x_n)$ introduces an onto map $f:\mathbf{F}_{S} \to \mathbf{F}_{S_{\text{fin}}}$ by following:
\[
f(\text{conv}(X))=\pi(\text{conv}(X)).
\]
We now prove that $f$ is a $p$-morphism. If $\text{conv}(X_0) \subseteq \text{conv}(X_1)$, then $f(\text{conv}(X_0))=\pi(\text{conv}(X_0))\subseteq\pi(\text{conv}(X_1))=f(\text{conv}(X_1))$. If $f(\text{conv}(X))=\pi (\text{conv}(X))\subseteq \text{conv}(Y)$, where $Y\subseteq S_{\text{fin}}$. Now consider the convex hull $\text{conv}( (Y\times\{1\})\cup X)$, it is obvious that $\text{conv}(X) \subseteq \text{conv}( (Y\times\{1\})\cup X)$, and $f(\text{conv}( (Y\times\{1\})\cup X))=\pi(\text{conv}( (Y\times\{1\})\cup X))=\text{conv}(Y)$. Therefore, $\mathbf{F}_{S_{\text{fin}}}$ is the $p$-morphic image of $\mathbf{F}_{S}$. Take the intersection of all logics of $\mathbf{F}_{S_{\text{fin}}}$ for arbitrary finite set $S_{\text{fin}}$, we obtain $\mathsf{KC}$ by Theorem~\ref{thm:Yop}. Consequently, $\mathsf{Y}_{\infty}^{\text{op}}\subseteq \bigcap_{S=S_{\text{fin}\times\mathbb{N}}} \mathsf{Log}(\mathbf{F}_S)\subseteq\mathsf{KC}$.

Putting everything together, we obtain that $\mathsf{Y}_{\infty}^{\text{op}}=\mathsf{KC}$.
\end{proof}

\subsection{Cantor set and $\mathsf{IPC}$}

\begin{definition}
The \emph{Cantor set} is the subset $\mathcal{C}_{\infty}\subset[0,1]$ obtained by iteratively deleting the open middle third of each remaining interval. More precisely, define $\mathcal{C}_0=[0,1]$ and $\mathcal{C}_{n+1}=\frac{1}{3}\mathcal{C}_n\cup \left(\frac{2}{3}+\frac{1}{3}\mathcal{C}_n\right)$, for $n\geq 0$.
Then $\mathcal{C}_{\infty}:=\text{lim}_{n\to \infty} \mathcal{C}_n =\bigcap_{n=0}^{\infty} \mathcal{C}_n$.
The open intervals removed during the $n$-th trisection step are called the \emph{$n$-th level intervals}.
\end{definition}
The Cantor set is compact, perfect, totally disconnected, uncountable, self-similar, and has Lebesgue measure zero. The following proof mainly relies on the removed intervals provided by the perfectness, disconnectedness, and self-similarity of the Cantor set $\mathcal{C}_{\infty}$. 

\begin{theorem}
$\mathsf{Y}_{\infty}=\mathsf{IPC}$.
\label{thm:YinfIPC}
\end{theorem}
\begin{proof}
Let $S=[0,1]\cap \mathbb{Q}$. We aim to prove that for any finite rooted frame $\mathbf{F}$, it is a $p$-morphic image of $\mathbf{F}_{S}$ by mathematical induction on the cardinality of the points of $\mathbf{F}$. 

If $\mathbf{F}$ is a single point, then the existence of $p$-morphism is obvious. Assume that $\mathbf{F}$ is not a singleton, it has a maximal point $a$, its root $r$ has $n$ immediate successors $s_0, s_1, \ldots, s_{n-1}$, then each $s_i\uparrow$ is a $p$-morphic image of $\mathbf{F}_{S}$ via $f_i$ by induction hypothesis. For any $m$-th level interval $X$, its convex hull is a closed interval $[x,y]$, where $x,y\in \mathbb{Q}$. The affine map $g(t):=x+(y-x)t$ is a bijection from $S$ to $S\cap [x,y]$, and it introduce an isomorphism $\mathbf{F}_S \cong [x,y]\uparrow$. Assume that $m\equiv i (\text{mod } n)$, then $s_i\uparrow$ in $\mathbf{F}$ is a $p$-morphic image of $[x,y]\uparrow$ in $\mathbf{F}_S$. The choice is well-defined since the convex hulls of those removed intervals in Cantor set are pairwise disjoint. Now we construct the map $f$ from $\mathbf{F}_S$ to $\mathbf{F}$ by following, for any convex hull $Y$ in $S$,
$$f(Y)=
\begin{cases}
f_i(Y), & \text{if $Y$ is contained in a $m$-th level interval and } m\equiv i (\text{mod } n)  ,\\
a, & \text{if $Y$ is a singleton and does not belong to any removed interval},\\
r, & \text{otherwise}.
\end{cases}$$ 

The forth condition for $f$ is obvious. For the back condition, suppose $f(Y)\leq y$, if $Y$ is contained in some removed interval, then the back condition follows the induction hypothesis. If $f(Y)=a$, then the case is trivial. The remaining case is $f(Y)=r$. Recall the property of Cantor set, for any open interval $X$, if $\mathcal{C}_{\infty}\cap X\neq \emptyset$, then for any $0\leq i<n$, there is an $m$-th level interval contained in $X$ such that $m\equiv i (\text{mod } n)$. If $\mathcal{C}_{\infty}\cap \text{int}(Y)=\emptyset$, then the connected $\text{int}(Y)$ belongs to a removed interval, and so $Y$ is contained in this removed interval, a contradiction. Thus $\mathcal{C}_{\infty}\cap \text{int}(Y)\neq\emptyset$. Since $r<y$, then there is some $s_i\leq y$ for $0\leq i<n$. There is an $X$ with its convex hull is contained in $Y$ and $X$ is an $m$-th level interval, $m\equiv i (\text{mod } n)$. So there exists a convex hull $Y_1$ contained in $\text{conv}(X)$ with $f(Y_1)=f_i(Y_1)=y$ and $Y_1\subseteq \text{conv}(X) \subseteq Y$. Therefore, $f$ is a $p$-morphism from $\mathbf{F}_S$ to $\mathbf{F}$. 

Since every finite rooted frame is a $p$-morphic image of $\mathbf{F}_S$, then $\mathsf{Log}(\mathbf{F}_S)\subseteq \mathsf{IPC}$. Together with $\mathbf{F}_S\vDash \mathsf{IPC}$, we have $\mathsf{Log}(\mathbf{F}_S)=\mathsf{IPC}$ and so $\mathsf{Y}_{\infty}=\mathsf{IPC}$.
\end{proof}

\section{Future work}
Two conjectures remain open. The first concerns the decidability of $\mathsf{Y}$. We proved that $\mathsf{Y}\in\Pi^0_1$ and that, for each $n\geq 1$, the logic $\mathsf{Y}^{\leq n}$ is decidable. However, these results do not settle the decidability of $\mathsf{Y}$ itself. We conjecture that $\mathsf{Y}$ is undecidable.
The second conjecture relates the entire interval $[\mathsf{Y},\mathsf{Med}]$. We proved that every logic in this interval is not finitely axiomatizable, and that every logic in $[\mathsf{Y}',\mathsf{Med}]$ is undecidable. We conjecture that every logic $\mathsf{M}$ satisfying $\mathsf{Y}\subseteq\mathsf{M}\subseteq\mathsf{Med}$ is undecidable. In particular, it remains to establish undecidability for those logics in this interval that do not contain $\mathsf{Y}'$, including $\mathsf{Y}$ itself.

\section*{Acknowledgements}
The authors would like to thank Johan van Benthem, Nick Bezhanishvili, and Martin Stokhof for insightful discussions.

None of the mathematical arguments, proofs presented in this paper were generated or developed using any AI technology. All such arguments, proofs were independently conceived and produced by the authors. 
The only AI-generated mathematical result used in the paper is the undecidability of Medvedev logic, which is used solely in the proofs of Theorems \ref{thm:Xisundecidable} and \ref{thm:Y'Medundecidable}.



\bibliographystyle{plainnat}
\bibliography{reference}

\end{document}